# Steering of beam trajectory by distorted photonic crystals

Kanji Nanjyo[1], Hitoshi Kitagawa[1], Daniel Headland[2], Masayuki Fujita[2**], and Kyoko Kitamura[1, 3*]

[1] *Department of Electronics, Kyoto Institute of Technology, Matsugasaki, Sakyo-ku, Kyoto 606-8585, Japan*

[2] *Graduate School of Engineering Science, Osaka University, 1-3 Machikaneyama, Toyonaka, Osaka 560-8531, Japan.*

[3] *Japan Science and Technology Agency, Precursory Research for Embryonic Science and Technology, 4-1-8 Honcho, Kawaguchi, Saitama, 332-0012, Japan*

*kyoko@kit.ac.jp, **fujita@ee.es.osaka-u.ac.jp

**Abstract:**

Electromagnetic waves follow linear paths in homogenous index media, with the exception of band edges. In this study, we introduced spatially distorted photonic crystals (D-PCs) that are capable of beam-steering light waves, even when a homogeneous refractive index is maintained. We analyzed their equi-frequency contours to investigate the correspondence between the direction of distortion and the direction of the group velocity vector in the D-PC. Thereafter, we experimentally verified the beam-steering phenomenon in the terahertz range using an all-silicon D-PC. Our structures serve as on-chip beam trajectory control without the need for any specially-engineered materials, using only lattice distortion.

# 1. Introduction

The propagation of electromagnetic (EM) waves in transparent media is controlled by the refractive index, and hence, a given desired beam trajectory may be obtained by engineering a spatially-varying refractive index. In practice, this may be achieved using metamaterials, which employ subwavelength resonators in order to artificially engineer both electro- and magnetic-permeability for arbitrary refractive index. This has previously enabled precise control of the propagation of EM waves, such as negative refraction [1], invisibility cloaks [2], beam steering, and birefringence [3–5]. However, metamaterials typically require lossy materials, such as metals, as well as narrowband resonance effects, in order to control permeability. Photonic crystals (PCs) are a different type of artificial material that are constructed by periodically arranging two or more dielectric media with different relative permittivities. PCs have an exotic optical property, i.e., a photonic band structure that corresponds to their periodicity. In contrast to a uniform medium, the group velocity of light waves propagating in a PC is defined by its photonic band structure, and this can be exploited for novel effects. For instance, the dispersion relation in the vicinity of photonic band edges may be tailored in order to produce phenomena such as negative refraction [6] and the superprism effect [7–10]. The latter is a phenomenon wherein the refractive angle changes rapidly according to variations in the frequency and incident angle due to an anisotropic dispersion relation. However, the dispersion relation is isotropic in the low-frequency region, similar to the relation in a uniform medium.

Although it is well known that conventional PCs exhibit a periodic structure [11], there is also interest in PC structures for which the periodicity is modified. For example, to realize high-$Q$ nanocavities, the $Q$-factor of PCs has been enhanced by shifting the lattice point positions [12–14]. In recent studies on laser beam steering by modulated PC lasers, two-dimensional resonant states are diffracted in any direction by modulated lattice point positions [15–17]. Note that, in these studies, the lattice point positions were moved within a range that was smaller than the PC lattice constant. However, the light propagation in a PC that exploits periodicity-breaking has not yet been fully explored.

In an electron system, gravitational effects can be generated by the deformation of crystals in the lower energy (or frequency) region [18]. In a photon system, light propagation has been shown to exhibit a curved trajectory due to a structure in which the lattice constant gradually expands in the uniaxial direction owing to the pseudo-magnetic field generated by the Dirac point, which is based on the topological notion of strained photonic honeycomb lattices [19–21]. Recently, we defined a PC that exhibits adiabatic change in the lattice constant as a distorted-PC (D-PC) and discussed the propagation of light waves from the viewpoint of differential geometry [22]. In that work, the solutions of the geodesic equation revealed that the light trajectory curved in the low-frequency range for a D-PC and for a comparable (non-distorted) PC, where light propagation is linear. This manner of bending can be referred to as pseudo-gravity caused by lattice distortion.

In this paper, we aim to explain such effects from the viewpoint of the group velocities and to experimentally demonstrate steering of the beam's in-plane trajectory. This paper is organized as follows: In Section 2, the light propagation in the D-PC is analyzed numerically using a two-dimensional finite-difference time-domain (2D-FDTD) method, in order to show that lattice distortion produces in-plane beam bending. In Section 3, we present the D-PCs as a series of PCs with varying rectangular lattice geometry and analyze their equi-frequency contours (EFCs) to investigate the correspondence between the direction of distortion and the direction of the group velocity vector. In Section 4, we experimentally verify this beam-steering phenomenon in the terahertz range with a D-PC slab that is implemented with a microscale through-hole lattice in an intrinsic silicon slab. We draw conclusions in Section 5.

## 2. Light propagation in the distorted photonic crystals

*2.1 Introducing distortion to a PC*

The square-lattice PC has lattice points represented by $x^{(m)} = ma$ and $y^{(n)} = na$, where $a$ is the lattice constant, and $m$ and $n$ are the numbers of lattice points in the $x$- and $y$-directions, respectively. Here, we discuss a unidirectional D-PC for which the lattice point position is distorted by

$$\begin{aligned} x^{(m)} &= ma, \\ y^{(n)} &= \begin{cases} na + n^2 a\beta & (n > 0) \\ na & (n < 0) \end{cases}, \end{aligned} \tag{1}$$

where $\beta$ is the distortion coefficient, which represents the magnitude of lattice distortion. To maintain the Bloch state, $\beta$ should be sufficiently smaller than $a$. Eq. (1) demonstrates that the interval of each lattice position increases gradually in proportion to the number of lattice points in the $y$-direction, as shown in Fig. 1 (a).

To investigate the propagation of light in this medium, we numerically analyzed the propagation of a Gaussian beam ($H$-mode: $H_z$-polarization, where the magnetic field is oriented perpendicular to the slab ($xy$-) plane) in both an air-hole and a dielectric rod D-PC. A schematic of the simulation model is shown in Fig. 1 (b). The parameters of the D-PCs are as follows: lattice point radius $r^{(0)} = 0.4a^{(0)}$, where $a^{(0)}$ is the lattice constant and dielectric constant: $\varepsilon_1 = 12.25$ (or 1), $\varepsilon_2 = 1$ (or 12.25). We input a Gaussian beam at an incident angle of 45° from the $x$-axis (Γ-M direction). To avoid excessive beam expansion, we set the input beam width to $w_0 = 2\lambda$, where $\lambda$ is the wavelength in free space. In view of the photonic band structure of the square lattice shown in Fig. 1 (c), we set the normalized frequency of the beam to 0.1 ($= a^{(0)}/\lambda$). Thus, the effects of peculiar points, such as the Dirac point and band-edge, can be eliminated, and it ensures that the beam has group velocity in the Γ-M direction. Moreover, we set the mesh size to $9\lambda/800$ with a dielectric constant volume average to express the distortion.

*2.2 Simulation results*

Figures 2(a) and (b) show the simulation results when the distortion coefficient is set to $\beta = 0.005$. As shown in Fig. 2, the light wave in the air-hole D-PC follows a straight path through the structure (Fig. 2 (a)), and the light wave in the D-PC propagates along a curved trajectory through the structure (Fig. 2 (b)). This discrepancy may be understood in terms of the effective index profile of both structures. For the air-hole D-PC, the effective refractive index gradually decreases in the +*y*-direction, as shown in Fig. 2 (c), and for the dielectric rod D-PC, the effective refractive index increases in the +*y*-direction, as shown in Fig. 2 (d). If the light propagation were simply to follows Snell's law, without the influence of PC effects, then the trajectory of light propagation in an air-hole D-PC should be bent toward the +*y*-axis, and the trajectory of light propagation in a dielectric-rod D-PC should be bent toward the +*x*-axis. However, since the air-hole D-PC results show straight light propagation, we postulate that the lattice distortion causes the light to bend in the opposite direction to the effective refractive index, as represented by the yellow arrows in Fig. 2 (c); hence, the two phenomena essentially cancel each other out. For the dielectric rod D-PC, we postulate that the distortion causes the bending of the light propagation in the same direction as the effective refractive index, as shown in Fig. 2 (d), and hence, light follows a curved path.

*2.3 Uniform-index PCs*

To isolate the effect of lattice distortion, we compensate for change in pitch by modifying the hole radius, as follows:

$$r^{(n)} = r^{(0)}\sqrt{1+2n\beta}. \quad (n>0), \qquad (2)$$

where $r^{(0)}$ is the lattice point radius of the square-lattice PC. In this manner, refraction effects due to variation in index are avoided. The simulation results of light propagation in the adaptive air-hole and dielectric rod D-PC are shown in Figs. 3 (a) and (b), respectively. The figures show that the light wave propagates with a similar degree of bending in both cases.

In addition, the curvature of the light propagation is proportional to the variation in the distortion coefficient, as shown in Fig. 4. If the curvature is defined as $\Delta\theta = \theta_0 - \theta_n$, where $\theta_0$ is the incident angle and $\theta_n$ is the angle at $x = na^{(0)}$, then $\Delta\theta = 14.5°$, $11.3°$, and $7.4°$ at $n = 60$ under the condition of $\beta = 0.007$, 0.005, and 0.003, respectively.

3.     **Analysis of equi-frequency contour**

*3.1 Effect of the lattice distortion*

Here, we analyze the propagation of light in the D-PC in terms of the group velocity vector and its variation owing to lattice distortion. The group velocity vector is the slope of the dispersion relation of the PC medium and is the normal vector at the intersection of the wave number vector and EFC. First, we approximate the D-PC as a series of uniform PCs with different lattice constants: a square-lattice PC (at $n = 0$) and rectangular PCs (at $n = 20, 40$), as illustrated in Fig. 5 (a). The EFC in each region under this approximation is calculated using the PWE method, as shown in Figs. 5 (b)-(d). Using these EFCs, we analyze the wave vector $k$ for light propagation in PC1 at a normalized frequency of 0.1 in the 45° direction from the $x$-axis. The interface component $k_x$ of the wave vector is conserved from the phase-matching conditions at the junction plane of the two media. Therefore, the wave vector in PC2 can be represented as $k'$. Consequently, the angle $\theta$ of the group velocity vector changes slightly from $\theta_1 = 45°$ in PC1 to $\theta_2 = 41°$ in PC2. In PC2 to PC3, the group velocity vector also slightly changed from $\theta_2 = 41°$ to $\theta_3 = 34°$. At the interface between two types of PCs with the same effective refractive index, the light wave is slightly bent owing to the change in shape of the EFC caused by lattice distortion. Therefore, in the D-PC, a light ray undergoes continuous minor changes of the direction of group velocity vector, resulting in the bending of the trajectory of light propagation. Although the change in shape of the EFC in *H*-mode is distinct, for *E*-mode ($E_z$ polarization, where the electric field is oriented perpendicular to the slab ($xy$-) plane) they are almost similar, as shown in Figs. 5 (e)-(g). The trajectory bending is quite small. This effect of the lattice distortion can be considered analogous to gravity within the context of distorted spacetime under the theory of general relativity [22].

*3.2 Light propagation with suppressed beam divergence*

In addition to steering the trajectory of the beam, it is possible to achieve light propagation with suppressed beam divergence by tailoring the shape of the EFC [23]. For a normalized frequency of 0.1, the shape of the EFC is curvilinear and almost elliptical. Therefore, even with a small wavenumber dispersion, the beam divergence is large. For a normalized frequency of 0.2, the shape of the EFC against wave vector *k* is almost flat, as shown Fig. 6 (a). Thus, as shown in Fig. 6 (b), the beam divergence is small, and collimated light propagation can be achieved.

## 4. Experimental validation

To experimentally verify the beam-steering phenomenon in a D-PC, we fabricated an air-hole D-PC in an intrinsic silicon slab. An air-hole structure is favored over dielectric rods for structural integrity reasons; for example, a dielectric slab that is perforated with air holes can be self-supporting [24–29], but rod arrays will require a support structure. To ensure a homogeneous refractive index, the radii of the air holes must be changed to compensate for the variation in the interval between the air holes, as described in Section 2.3. Consequently, the edges of adjacent air holes come closer and finally connect with increasing *n*. Considering the fabrication precision for such small intervals, we prepared samples with microscale feature sizes, corresponding to the terahertz regime.

Two samples were devised to support the experimental verification: a uniform square-lattice PC and a D-PC. Figure 7 shows the design of these sample structures, as well as micrographs at specific positions that are indicated in the design. The square-lattice PC and the D-PC were arranged with one input (port A) and two output ports (ports B and C) as shown in Figs. 7 (a) and (b). Both PCs are designed with $a^{(0)} = 200$ μm and $r^{(0)} = 80$ μm ($= 0.4\, a^{(0)}$), and for the D-PC, $\beta = 0.001845$. This value is selected considering two issues. First, $\beta$ must be maximized to accentuate the steering effect while avoiding physical connection between adjacent air holes, which would compromise the structural integrity of the sample. Figures 7 (c) and (e) show that the air holes of the D-PC are larger than those of the square-lattice PC, but the structure remains self-supporting despite the holes' physical connection. Second, plane wave excitation is required to avoid strong diffraction within the PC slab, and all ports are coupled to GRIN lenses that interface between a diffraction-resistant in-slab collimated beam and a channel waveguide [30, 31]. The physical structure and performance of the lenses have previously been discussed in detail [31]. The lens that is coupled to port A launches a plane wave into the D-PC medium, and ports B and C subsequently collect the power following transit through the sample and convey it to a receiver. The beam width of plane wave generated from port A is 8 mm when the input electromagnetic wave frequency is 0.3 THz ($H$-mode, normalized frequency becomes 0.2 for this structure). For the square-lattice PC, the plane wave propagates linearly and will subsequently be divided equally between ports B and C. For the D-PC, the plane wave will bend in the direction of port C, and hence, there should be a discrepancy in the intensity received by ports B and C.

Figures 8 (a) shows the experimental setup employed to probe the transmission between port pairs A-B and A-C of both the square-lattice PC and the D-PC sample. An all-electronic measurement system was employed for this purpose. Terahertz waves are generated by up-conversion of a mm-wave signal of ~32–42 GHz by a ×9 frequency multiplier and are subsequently conveyed to a hollow rectangular metallic waveguide with inner conductor dimensions of 711 μm × 356 μm. A linear tapered spike is situated at the termination of the dielectric channel waveguide that interfaces with the integrated lens at port A. This spike is inserted directly into the hollow waveguide to provide broadband index matching and facilitate efficient flow of power into the silicon sample. An alignment between the silicon sample and the hollow waveguide was performed manually using precise micrometer stages, and the sample was held with ordinary tweezers during probing. A detector is interfaced to the sample via a second hollow metallic waveguide, which is connected to either ports B or C. This down-converts the received terahertz power with a mixer and a ×36 multiplier, thereby yielding signals that may be displayed by a microwave spectrum analyzer. For both the square-lattice PC and D-PC samples, ports B and C are probed in this manner. An overview image of the experiment is shown in Fig. 8 (b) and the experimental results are shown in Fig. 8 (c). For a square-lattice PC, near-identical intensity is detected across the 0.285–0.380 THz ($\omega a^{(0)}/2\pi c$ = 0.19–0.26) measurement range, as designed; hence, the ratio of C to B is centered upon 0 dB. The D-PC shows strong skewness toward port C for all frequencies lower than 0.375 THz ($\omega a^{(0)}/2\pi c$ < 0.25). The D-PC exhibits a ratio of ~20 dB in the received intensity between ports C and B in the region of 0.300–0.330 THz ($\omega a^{(0)}/2\pi c$ = 0.20–0.22). Therefore, the electromagnetic wave is indeed curved within the D-PC, as designed, and this validates the principle of beam steering in D-PCs. Note that, the decrease in the ratio for frequencies higher than 0.375 THz ($\omega a^{(0)}/2\pi c$ < 0.25) is caused by a change in the shape of the EFC owing to the proximity to the band-edge.

5.  **Conclusion**

In this study, we introduced the concept of D-PCs that are capable of beam-steering light waves, even when a homogeneous refractive index is maintained. We employed FDTD simulations and demonstrated that the trajectory of light propagation in the D-PC is curved for both air-hole and dielectric rod D-PCs. The physics underlying the trajectory bending was explained by analyzing the EFC of two types of PCs. Based on the analysis, we confirmed that the shape of the EFC is slightly changed by providing a lattice distortion to the PC, and that the direction of the group velocity vector is also slightly changed at the interface between PCs. Furthermore, we experimentally confirmed the beam-steering phenomenon with an all-silicon microscale D-PC, which we probed using terahertz waves. We also confirmed that the trajectory bending is quite small for the E-mode. Therefore, the D-PC exhibits the characteristics of structural optical birefringence. Pseudo-gravity caused by lattice distortion of homogeneous refractive media will open up new approaches to achieve on-chip trajectory control of light propagation in PCs. Such in-plane trajectory curvature has potential to apply to spatial phase control of beam by combining with vertical diffraction. This work has achieved a form of light trajectory control that can be considered analogous to the distortion of spacetime under the theory of general relativity, and provides a pathway to realize novel devices that exploit gravitational effects such as gravitational lens and gravitational wave.

**Acknowledgement.**

This work was partly supported by the Precursory Research for Embryonic Science and Technology (PRESTO) program, Japan Science and Technology Agency (JST) (Grant No. JP 20345471), Core Research for Evolutional Science and Technology (CREST) program, Japan Science and Technology Agency (JST) (Grant No. JPMJCR1534 and JPMJCR21C4), and a Grant-in-Aid for Scientific Research, the Ministry of Education, Culture, Sports, Science and Technology of Japan (MEXT) (Grant No. 20H01064).

**References.**

[1] R. A. Shelby, D. R. Smith, and S. Schultz, "Experimental verification of a negative index of refraction," Science **292**(5514), 77–79 (2001).

[2] J. B. Pendry, D. Schurig, and D. R. Smith, "Controlling electromagnetic fields," Science **312**(5781), 1780–1782 (2006).

[3] C. Pfeiffer, N. K. Emani, A. M. Shaltout, A. Boltasseva, V. M. Shalaev, and A. Grbic, "Efficient light bending with isotropic metamaterial Huygens' surfaces," Nano Lett. **14**(5), 2491–2497 (2014).

[4] Y. Li, and Q. Zhu, "Broadband birefringent metamaterial lens with bi-functional high-gain radiation and deflection properties," Opt. Express **26**(13), 16265–16276 (2018).

[5] S. Maurya, M. Nyman, M. Kaivola, and A. Shevchenko, "Highly birefringent metamaterial structure as a tunable partial polarizer," Opt. Express **27**(19), 27335–27344 (2019).

[6] M. Notomi, "Theory of light propagation in strongly modulated photonic crystals: Refractionlike behavior in the vicinity of the photonic band gap," Phys. Rev. B **62**(16), 10696–10705 (2000).

[7] S. Y. Lin, V. M. Hietala, L. Wang, and E. D. Jones, "Highly dispersive photonic band-gap prism," Opt. Lett. **21**(21), 1771–1773 (1996).

[8] H. Kosaka, T. Kawashima, A. Tomita, M. Notomi, T. Tamamura, T. Sato, and S. Kawakami, "Superprism phenomena in photonic crystals," Phys. Rev. B **58**(16), R10096–R10099 (1998).

[9] T. Baba, and M. Nakamura, "Photonic crystal light deflection devices using the superprism effect," IEEE J. Quantum Electron. **38**(7), 909–914 (2002).

[10] J. Upham, B. Gao, L. O'Faolain, Z. Shi, S. A. Schulz, and R. W. Boyd, "Realization of a flat-band superprism on-chip from parallelogram lattice photonic crystals," Opt. Lett. **43**(20), 4981–4984 (2018).

[11] M. Fujita, S. Takahashi, Y. Tanaka, T. Asano, and S. Noda, "Simultaneous inhibition and redistribution of

spontaneous light emission in photonic crystal," Science **308**, 1296–1298 (2005).

[12] Y. Akahane, T. Asano, B. S. Song, and S. Noda, "High-Q photonic nanocavity in a two-dimensional photonic crystal," Nature **425**(6961), 944–947 (2003).

[13] B.-S. Song, S. Noda, T. Asano, and Y. Akahane, "Ultra-high-Q photonic double-heterostructure nanocavity," Nature Mater. **4**(3), 207–210 (2005).

[14] H. Sekoguchi, Y. Takahashi, T. Asano, and S. Noda, "Photonic crystal nanocavity with a Q-factor of ~9 million," Opt. Express **22**(1), 916–924 (2014).

[15] S. Noda, K. Kitamura, T. Okino, D. Yasuda, and Y. Tanaka, "Photonic-crystal surface-emitting lasers: Review and introduction of modulated-photonic crystals," IEEE J. Quantum Electron. **23**(6), 1–7 (2017).

[16] K. Kitamura, T. Okino, D. Yasuda, and S. Noda, "Polarization control by modulated photonic-crystal lasers," Opt. Lett. **44**(19)**,** 4718–4720 (2019).

[17] R. Sakata, K. Ishizaki, M. De Zoysa, S. Fukuhara, T. Inoue, Y. Tanaka, K.

Iwata, R. Hatsuda, M. Yoshida, J. Gelleta, and S. Noda, "Dually modulated photonic crystals enabling high-power high-beam-quality two-dimensional beam scanning lasers," Nat. Commun. **11**(1), 3487 (2020).

[18] L. Dong, and Q. Niu, "Geometrodynamics of electrons in a crystal under position and time-dependent deformation," Phys. Rev. B **98**(11) 115162 (2018).

[19] F. Deng, Y. Li, Y. Sun, X. Wang, Z. Guo, Y. Shi, H. Jiang, K. Chang, and H. Chen, "Valley-dependent beams controlled by pseudomagnetic field in distorted photonic graphene," Opt. Lett. **40**(14), 3380–3383 (2015).

[20] H. Schomerus, and N. Y. Halpern, "Parity anomaly and Landau-level lasing in strained photonic honeycomb lattices," Phys. Rev. Lett. **110**(1), 013903 (2013).

[21] M. C. Rechtsman, J. M. Zeuner, A. Tünnermann, S. Nolte, M. Segev, and A. Szameit, "Strain-induced pseudomagnetic field and photonic Landau levels in dielectric structures," Nature Photon. **7**(2), 153–158 (2013).

[22] H. Kitagawa, K. Nanjyo, and K. Kitamura, "Effective field theory for distorted photonic crystals," Phys. Rev. A **103**(6) 063506 (2021).

[23] L. Zhang, Q. Zhan, J. Zhang, and Y. Cui, "Diffraction inhibition in two-dimensional photonic crystals," Opt. Lett. **36**(5), 651–653 (2011).

[24] R. Kakimi, M. Fujita, M. Nagai, M. Ashida, and T. Nagatsuma, "Capture of a terahertz wave in a photonic crystal slab," Nature Photon. **8**(8), 657–663 (2014).

[25] K. Tsuruda, M. Fujita, and T. Nagatsuma, "Extremely low-loss terahertz

waveguide based on silicon photonic crystal slab," Opt. Express **23**(25), 31977–31990 (2015).

[26] D. Headland, X. Yu, M. Fujita, and T. Nagatsuma, "Near-field out-of-plane coupling between terahertz photonic crystal waveguides," Optica **6**(8), 1002–1011 (2019).

[27] D. Headland, W. Withayachumnankul, X. Yu, M. Fujita, and T. Nagatsuma, "Unclad microphotonics for terahertz waveguides and systems," J. Lightwave Technol. **38**(24), 6853–6862 (2020).

[28] Y. Yang, Y. Yamagami, X. Yu, P. Pitchappa, J. Webber, B. Zhang, M. Fujita, T. Nagatsuma, and R. Singh, "Terahertz topological photonics for on-chip communication," Nature Photon. **14**(7), 446–451 (2020).

[29] D. Headland, W. Withayachumnankul, M. Fujita, and T. Nagatsuma, "Gratingless integrated tunneling multiplexer for terahertz waves," Optica **8**(5), 621–629 (2021).

[30] D. Headland, M. Fujita, and T. Nagatsuma, "Half-Maxwell fisheye lens with photonic crystal waveguide for the integration of terahertz optics," Opt. Express **28**(2), 2366–2380 (2020).

[31] D. Headland, A. K. Klein, M. Fujita, and T. Nagatsuma, "Dielectric slot-coupled half-maxwell fisheye lens as octave-bandwidth beam expander for terahertz-range applications," APL Photon. **6**, 096104 (2021).

**Caption of Figures.**

Fig. 1. (a) Schematic of lattice point positions of the D-PC based on the square-lattice PC. The D-PC is accorded a lattice distortion in which the interval of lattice points gradually expands in the $+y$-direction. The $x$- and the $y$- axes are defined along the Γ-X direction of the square-lattice PC. (b) Simulation model diagram. The Gaussian beam is excited at the origin of square-lattice PC. Eight perfectly matched layers (PML) absorbing boundaries are employed. (c) Photonic band structure of an air-hole square-lattice PC calculated using the plane wave expansion (PWE) method.

Fig. 2. Light propagation in the D-PC with $\beta = 0.005$ ($H$-mode, the $z$-component of the magnetic field can be observed). (a) Simulation results for the air-hole D-PC and (b) dielectric rod D-PC. The axis scale is the length normalized by the lattice constant $a^{(0)}$. Diagonal dotted lines show the path that light would take in a homogeneous medium. Effects and directions that affect the light propagation in (c) air-hole D-PCs and (d) dielectric rod D-PCs.

Fig. 3. Light propagation in the adaptive D-PC with $\beta = 0.005$ ($H$-mode). (a) Simulation results for the air-hole D-PC and (b) dielectric rod D-PC. The light wave propagates with the same degree of bending in both results. The direction in which distortion effects light propagation in (c) air-hole D-PC and (d)

dielectric rod D-PC.

Fig. 4. Light propagation in the adaptive air-hole D-PC (*H*-mode). (a) $\beta = 0.007$, (b) $\beta = 0.005$, (c) $\beta = 0.003$. The curvature of the light propagation is proportional to the variation of the distortion coefficient.

Fig. 5. (a) Schematic of the D-PC approximated as a series of PCs: PC1 represents a square-lattice PC, PC2 represents a rectangular lattice PC with $a_y = 1.2a^{(0)}$ (corresponding D-PC at $n = 20$), and PC3 represents a rectangular lattice PC with $a_y = 1.4a^{(0)}$ (corresponding D-PC at $n = 40$). (b)-(d) Calculation results of EFC in *H*-mode for each PC3-1, respectively. (e)-(g) Calculation results of EFC in *E*-mode for each PC3-1, respectively. The vertical and horizontal axes of the EFC are normalized using the *x*-direction lattice constant $a^{(0)}$. In the figure, $k$, $k'$, and $k''$ represent the wave number vector of light waves propagating in each PC, and $v_g$, $v_g'$, and $v_g''$ represent the group velocity vector.

Fig. 6. (a) EFC of a rectangular PC corresponding to $n = 20$ in a D-PC with $\beta = 0.005$. (b) Light propagation in the Γ-M direction at a normalized frequency of 0.2.

Fig. 7. Design of square-lattice PC (a) and D-PC (b). Both samples have one input port (Port A) and two output ports (Port B, C) that use GRIN lenses

structure. (c)-(f) Micrographs of fabricated samples at the locations indicated in (a) and (b).

Fig. 8. (a) Schematic diagram of experimental setup. (b) Overview image of experimental setup. (b) Output signal ratio of C to B. Red indicates D-PC and blue indicates square-lattice PC. Dots indicate raw data and lines indicate average values of adjacent 15 raw data. The variation in these results is due to Fabry-Perot reflections at the input port and output ports.

**Figures.**

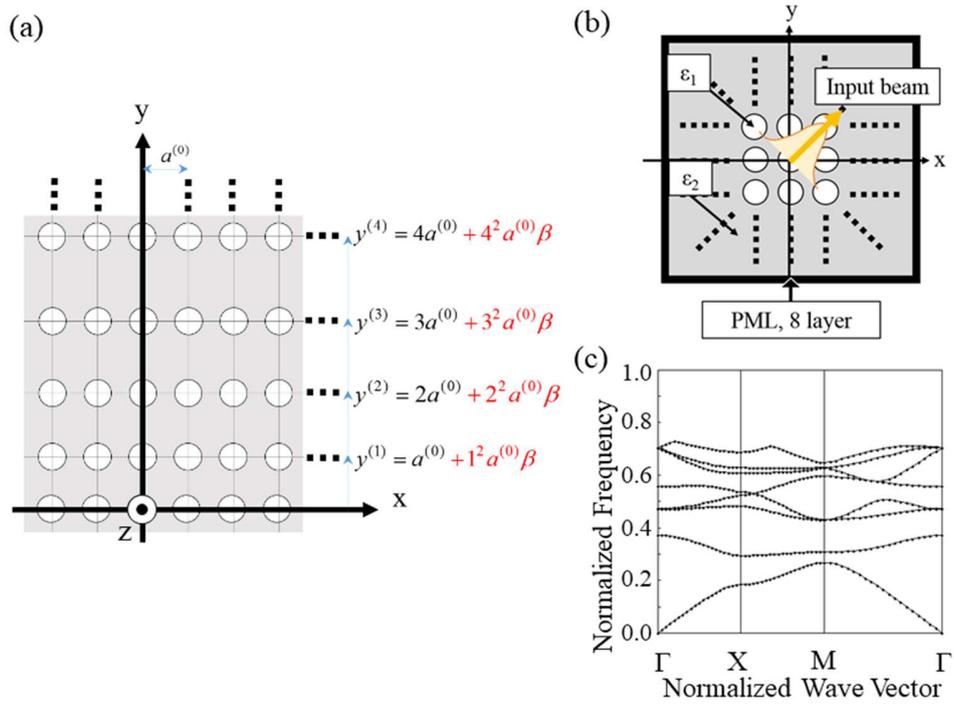

Fig.1

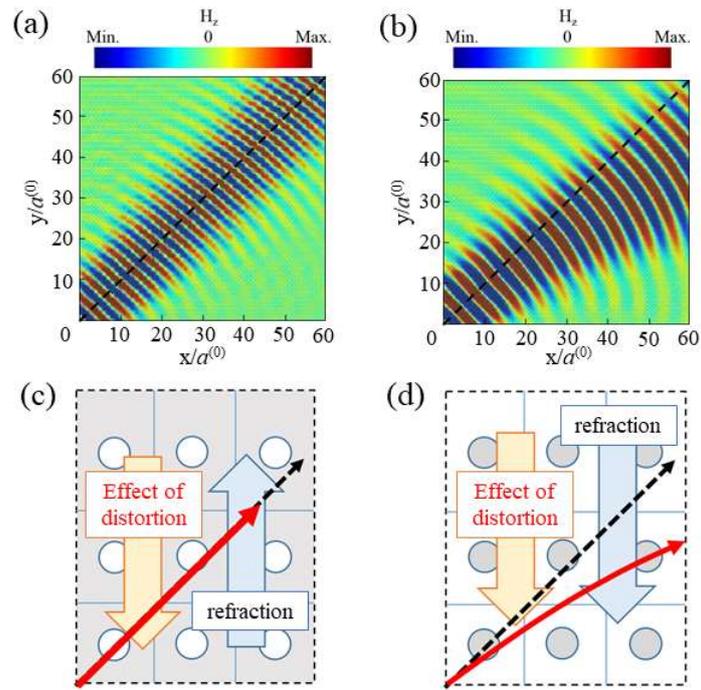

Fig.2

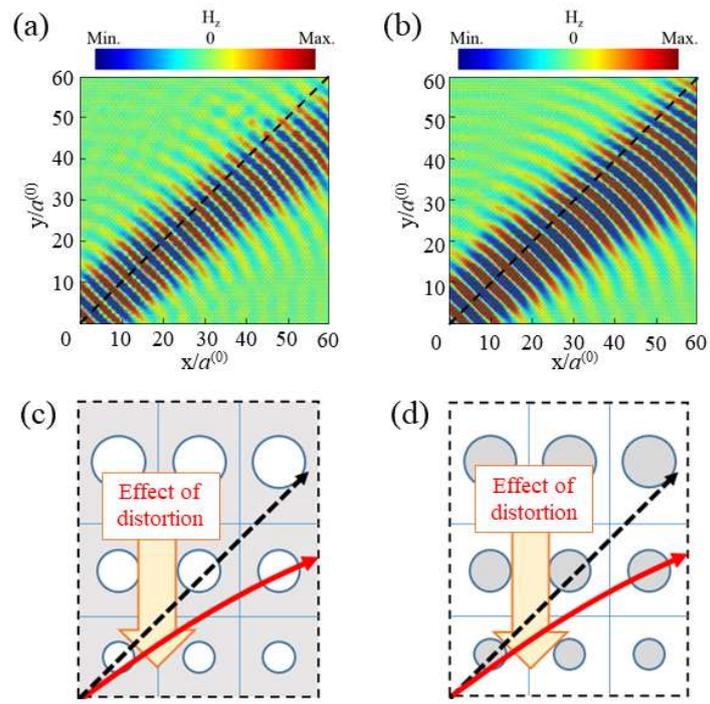

Fig.3

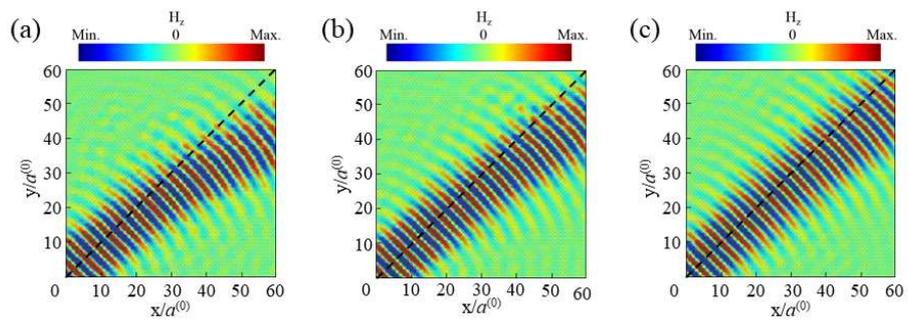

Fig.4

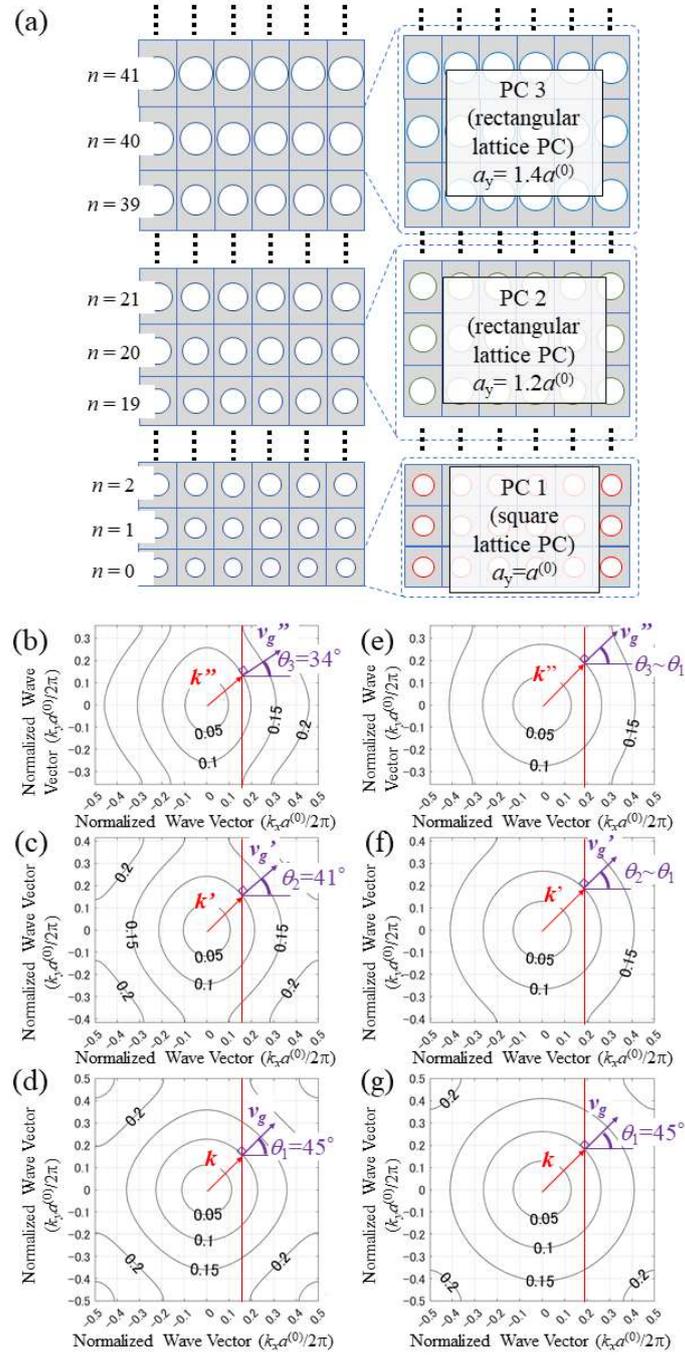

Fig. 5

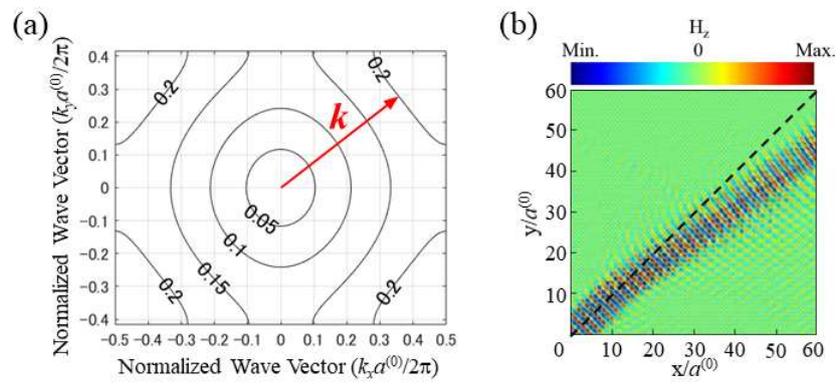

Fig.6

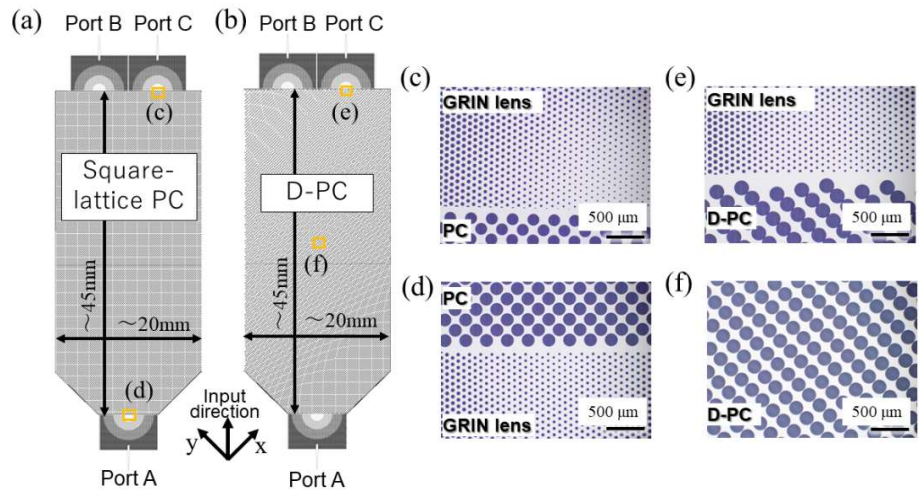

Fig. 7

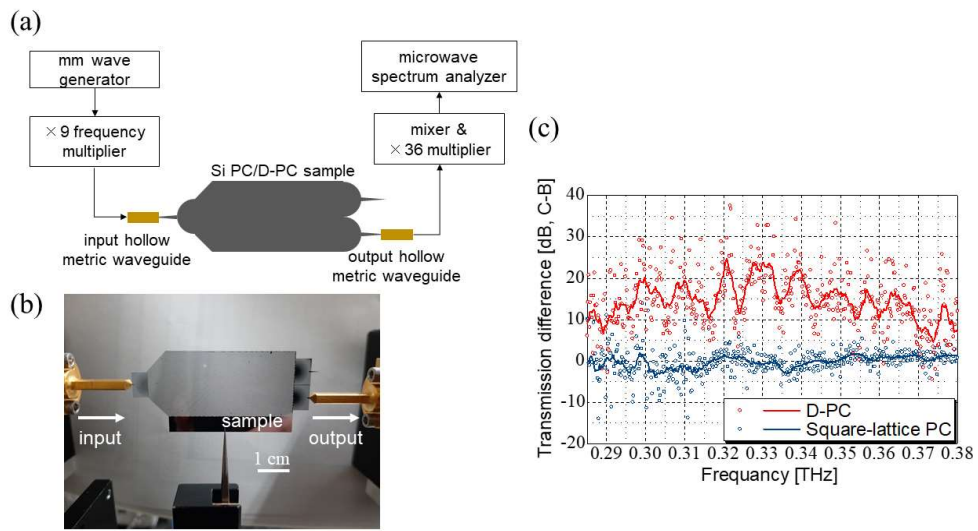

Fig.8